\documentclass[preprint]{aastex}

\shortauthors{Charbonneau et al.\ 2005}
\shorttitle{Transit Photometry of HD~149026b}

\begin{document}

% ------------------------------------------------------------------------
% New commands
%
\newcommand{\C}{\ensuremath{^{\circ}C\;}}
\newcommand{\el}{\ensuremath{e^-}}
\newcommand{\sqarcsec}{\ensuremath{\Box^{\prime\prime}}}
\newcommand{\sqarcdeg}{\ensuremath{\Box^{\circ}}}
\newcommand{\aduel}{\ensuremath{\lbrack ADU/\el \rbrack}}
\newcommand{\eladu}{\ensuremath{\rm {\lbrack \el/ADU \rbrack}}}
\newcommand{\adupixs}{\ensuremath{\rm ADU/(pix\, s)}}
                                                                                          
% -------------------------------------------------------------------------
%

\bibliographystyle{apj}

\title{Transit Photometry of the Core-Dominated Planet HD~149026b}

\author{David Charbonneau, Joshua N.\ Winn\altaffilmark{1}, David W.\ Latham, 
G\'asp\'ar Bakos\altaffilmark{1}, Emilio E.\ Falco,\\ Matthew J.\ Holman, 
Robert W.\ Noyes, and Bal\'azs Cs\'ak\altaffilmark{2}}
\affil{Harvard-Smithsonian Center for Astrophysics, 60 Garden Street, Cambridge, MA 02138;
dcharbonneau@cfa.harvard.edu}

\author{Gilbert A.\ Esquerdo and Mark E.\ Everett}
\affil{Planetary Science Institute, 1700 East Fort Lowell, Tucson, AZ 85719}

\and

\author{Francis T.\ O'Donovan}
\affil{California Institute of Technology, 1200 East California Blvd.,
Pasadena, CA 91125}

\altaffiltext{1}{Hubble Fellow.}
\altaffiltext{2}{SAO Predoctoral Fellow.}

\begin{abstract}
We report $g$, $V$, and $r$ photometric time series of HD~149026
spanning predicted times of transit of the Saturn-mass planetary companion,
which was recently discovered by Sato and collaborators. 
We present a joint analysis of our observations and the previously
reported photometry and radial velocities of the central star.  
We refine the estimate of the transit ephemeris to 
$T_c~[{\rm HJD}] = (2453527.87455^{+0.00085}_{-0.00091}) + (2.87598^{+0.00012}_{-0.00017}) \ N$.  
Assuming that the star has a radius 
of $1.45\pm 0.10~R_\odot$ and a mass of $1.30\pm 0.10~M_\odot$, we 
estimate the planet radius to be $0.726\pm 0.064\ R_{\rm Jup}$,
which implies a mean density of $1.07^{+0.42}_{-0.30}\ {\rm g \, cm^{-3}}$. 
This density is significantly greater than that predicted
for models which include the effects of stellar insolation
and for which the planet has only a small core of solid material.
Thus we confirm that this planet likely contains a large core,
and that the ratio of core mass to total planet mass is
more akin to that of Uranus and Neptune than that of either Jupiter or Saturn.
\end{abstract}


\section{Introduction}

Sato et al.\ (2005) recently presented the discovery of a planetary companion to the 
bright G0~{\sc iv} star HD~149026. The star exhibits a
time-variable Doppler shift that is consistent with a sinusoid of
amplitude $K=43$~m~s$^{-1}$ and period $P=2.9$~days, which would be
produced by the gravitational force from an orbiting planet with
$M_{\rm P}\sin i = 0.36$~$M_{\rm Jup}$. Furthermore, at the predicted
time of planet-star conjunction, the star's flux declines by 0.3\% in
the manner expected of an eclipse by a planet of radius $0.72$~$R_{\rm
Jup}$ (given an estimate of the stellar radius, 1.45~$R_\odot$, that
is based on the stellar parallax and effective temperature). Sato et al.\ (2005) observed
three such eclipses. This discovery is extraordinary for at least two reasons.

Firstly, the occurrence of eclipses admits this system into the elite
club of bright stars with detectable planetary transits. Of all the
previously-known transiting systems, only HD~209458 (Charbonneau et al.\ 2000;
Henry et al.\ 2000) and TrES-1 (Alonso et al.\ 2004; Sozzetti et al.\ 2005) have
parent stars brighter than $V=12$, and therefore only they are
amenable to a number of fascinating measurements requiring a very high
signal-to-noise ratio. Among these studies are (i) the search for satellites and
rings (Brown et al.\ 2001), (ii) the search for period variations due to
additional companions (Wittenmyer et al.\ 2005), (iii) the detection of
(or upper limits on)
atmospheric absorption features in transmission (Charbonneau et al.\ 2002; 
Brown, Libbrecht, \& Charbonneau 2002; Deming et al.\ 2005a), (iv) the 
characterization of the exosphere (Bundy \& Marcy 2000; Moutou et al.\ 2001, 2003; 
Vidal-Madjar et al.\ 2003, 2004; Winn et al.\ 2004; Narita et al.\ 2005),
(v) the measurement of the angle between the sky-projected orbit
normal and stellar rotation axis (Queloz et al.\ 2000; Winn et al.\ 2005), 
and (vi) the search for spectroscopic features near the
times of secondary eclipse (Richardson et al.\ 2003a, 2003b), 
and (vii) the direct detection of thermal 
emission from the planet (Charbonneau et al.\ 2005; Deming et al.\ 2005b).  
Charbonneau (2004) reviews these techniques
and related investigations.

Secondly, the planet is the smallest and least massive of the 8 
known transiting extrasolar planets\footnote{In addition to HD~209458b and TrES-1,
the OGLE photometric survey (Udalski et al.\ 2002, 2004)
and spectroscopic follow-up efforts
have located 5 such objects.  
Recent estimates of the planetary radii have been given by Bouchy et al.\ (2004), Holman et al.\ (2005), 
Konacki et al.\ (2003), Moutou et al.\ (2004), Pont et al.\ (2004), and references therein.}. 
This makes HD~149026b an
important test case for theories of planetary structure. Sato et al.\
(2005) argued that, once the effects of stellar insolation are included,
the small planetary radius implies that the planet has a
large and dense core.  In particular, 
assuming a core density $\rho_{c} = 5.5\ {\rm g \, cm^{-3}}$,
their models predict a prodigious core mass of 78~Earth masses, or 74\% of the total mass
of the planet. 
This, in turn, would
seemingly prove that the planet formed through core accretion, as
opposed to direct collapse through a gravitational instability.

A system of such importance should be independently confirmed, and the
determination of its basic parameters should be refined through
multiple observations.  With this as motivation, we performed
photometry of HD~149026 on two different nights when transits were
predicted by Sato et al.\ (2005). These observations and the data
reduction procedures are described in \S~2. The model that we used to
determine the system parameters is described in \S~3, and the results
are discussed in \S~4. Our data are available in digital form in the
electronic version of this article, and from the authors upon request.

\section{The Observations and Data Reduction}

\begin{boldmath}
\subsection{FLWO 1.2m $g$ and $r$ Photometry}
\end{boldmath}

We observed HD~149026 ($V=8.16$,
$B-V=0.56$) on UT~2005~June~6 and UT~2005~July~2 with the 48-inch 
(1.2m) telescope of the F.~L.~Whipple Observatory (FLWO) located at Mount Hopkins,
Arizona. We used Minicam, an optical CCD imager with two
$2048\times 4608$ chips.  In order to increase the duty cycle of the
observations, we employed $2\times 2$~binning, which reduced the readout 
and overhead time
to 20~s.  Each binned pixel subtends approximately $0\farcs6$ on the sky, giving an
effective field of view of about $10\arcmin \times 23\arcmin$ for each CCD.  
Fortunately, there exists a nearby object of similar brightness
and color (HD~149083; $V=8.05$, $B-V=0.40$, $\Delta\alpha$~$=5.1\arcmin$, $\Delta\delta = -17\arcmin$),
which we employed as an extinction calibrator.
We selected the telescope pointing so that both stars were imaged
simultaneously.  We defocused the telescope so that the full-width at 
half-maximum (FWHM) of a stellar image was typically 15
binned pixels (9$\arcsec$), and we used automatic guiding to ensure
that the centroid of the stellar images drifted no more than
3 binned pixels over the course of the night.  In addition to enabling longer integration times,
this served to mitigate the effects of pixel-to-pixel sensitivity variations
that were not perfectly corrected by our flat-fielding procedure.
On UT~2005~June~6, we gathered 5.5~hrs of SDSS $g$-band observations with typical
integration times of 8~s and a cadence of 28~s.  The conditions were photometric, and
the frames span an airmass from 1.01 to 1.74.  On UT~2005~July~2, we gathered 
4.4~hrs of SDSS $r$-band observations with integration times of 6~s and a median 
cadence of 26~s.  The field appeared to remain free of clouds for the duration of
the observations, which spanned an airmass range of 1.01 to 1.43,
although occasional patches of high cirrus could be seen in 
images from the MMT all-sky camera.

We converted the image time stamps to Heliocentric Julian Day (HJD) at mid-exposure.
The images were overscan-subtracted, trimmed, and divided by a
flat-field image. We performed aperture photometry of HD~149026 and
the comparison star HD~149083, using an aperture radius of 15 binned pixels
(9$\arcsec$) for the UT~2005~June~6 data, and an aperture radius of 20 binned
pixels (12$\arcsec$) for the UT~2005~July~2 data.  We subtracted the underlying
contribution from the sky for both the target and calibrator by estimating the counts in an annulus 
exterior to the photometric aperture.  The relative flux of HD~149026 was computed as the
ratio of the fluxes within the two apertures.  Normalization and residual extinction 
corrections are described in \S~3.

\begin{boldmath}
\subsection{TopHAT $V$ Photometry at FLWO}
\end{boldmath}

TopHAT is an automated telescope located on Mt.~Hopkins, Arizona,
which was designed to perform multi-color photometric follow-up of transiting
extrasolar planet candidates identified by the HAT network (Bakos et al.\ 2004).
Since TopHAT has not previously been described in the
literature, we digress briefly to outline the principal goals and features
of the instrument.

Wide-field transit surveys must contend with a large
rate of astrophysical false positives, which result from stellar systems that
contain an eclipsing binary and precisely mimic the single-color
photometric light curve of a Jupiter-sized planet transiting a Sun-like
star (Brown 2003; Charbonneau et al.\ 2004; Mandushev et al.\ 2005; Torres et al.\ 2004).
Although multi-epoch radial velocity follow-up is an effective
tool for identifying these false positives (e.g. Latham 2003),
instruments such as TopHAT and Sherlock
(Kotredes et al.\ 2004) can be fully-automated, and thus offer
a very efficient means of culling the bulk of such false positives.
TopHAT is a 0.26m diameter f/5 commercially-available
Baker Ritchey\--Chr\'etien telescope on an equatorial fork mount 
developed by Fornax Inc.
A 1$\fdg$25-square field of view is imaged onto a 2k$\times$2k 
Peltier-cooled, thinned CCD detector, yielding a pixel scale of 2$\farcs$2.
The time for image readout and associated overheads is 25~s.  
Well-focused images have a typical FWHM of 2~pixels. 
A two-slot filter-exchanger permits imaging in either $V$ or $I$.
The components are protected from inclement weather by an
automated asymmetric clamshell dome. 

We observed HD~149026 on UT 2005 July 2, the same night
as the FLWO 1.2m $r$ observations described above.
In order to extend the integration times and increase
the duty cycle of the observations, we broadened
the point spread function (PSF) by performing small, regular motions in RA and DEC
according to a prescribed pattern that was repeated during each
13~s integration (see Bakos et al.\ 2004 for details).  The 
resulting PSF had a FWHM of 3.5 pixels (7$\farcs$7).  We gathered
4.8~hrs of $V$ observations with a cadence of 68~s, spanning
an airmass range of 1.01 to 1.45. 

We converted the time stamps in the image headers to HJD at mid-exposure.
We calibrated the images by
subtracting the overscan bias and a scaled dark image, and dividing
by an average sky flat from which large outliers had been rejected.
We evaluated the centroids of the target and the three brightest
calibrators in each image.  For each star, we summed the flux 
within an aperture with a radius of 8~pixels (17$\farcs$6), and subtracted
a local sky estimate based on the median flux in an annulus
exterior to the photometric aperture.  We divided the resulting
time series for the target by the statistically-weighted average 
of the time series for the three calibrator stars.  The resulting relative flux time
series was then corrected for normalization and residual
extinction effects as described in the following section.

\section{The Model}

We attempted to fit simultaneously (i) the 3 photometric time series 
discussed above, (ii) the $(b+y)/2$ photometry of 3 transits presented 
by Sato et al.\ (2005), and (iii) the 7 radial velocities that were measured by Sato et
al.\ (2005) when the planet was not transiting. We did not attempt to
fit the 4 radial velocities measured during transits, which would
have required a model of the Rossiter-McLaughlin effect (Queloz et al.\
2002; Ohta, Taruya, \& Suto 2004; Winn et al.\ 2005).

We modeled the system as a circular Keplerian orbit. Following Sato et
al.\ (2005), we assumed the stellar mass ($M_{\rm S}$) to be
$1.30~M_\odot$ and the stellar radius\footnote{With more accurate
photometry and better time sampling of the ingress or egress, it would
be possible to solve for the stellar radius, rather than assuming a
certain value (see, for example, Brown et al.\ 2001, Winn et al.\
2005, Wittenmyer et al.\ 2005, and Holman et al.\ 2005). In this case,
we found that the stellar radius is not well determined by the
photometry. Rather, the constraint on the stellar radius based on
stellar parallax and spectral modeling is tighter.} ($R_{\rm S}$) to
be $1.45~R_\odot$. The free parameters were the planetary mass
($M_{\rm P}$), planetary radius ($R_{\rm P}$), orbital inclination
($i$), orbital period ($P$), central transit time ($T_c$), and the
heliocentric radial velocity of the center of mass ($\gamma$). We
included 2 free parameters for each of our 3 photometric time series:
an overall flux scaling $C$; and a residual extinction coefficient
($k$) to correct for differential extinction between the target star
and the comparison object, which have somewhat different colors. These
are defined such that the relative flux observed through an airmass
$X$ is $C\exp(-kX)$ times the true relative flux. The residual
extinction corrections were small but important at the millimagnitude
level; for example, in the $g$ band, we found $k\approx -2\times
10^{-3}$. The 3 time series presented by Sato et al.\ (2005) were
already corrected for airmass, so we allowed each of these to have
only an independent flux scaling.  Finally, we allowed the data from
each spectrograph (Keck/HIRES and Subaru/HDS) to have an independent
value of $\gamma$.

We computed the model radial velocity at each observed time as $\gamma
+ \Delta v_r$, where $\Delta v_r$ is the line-of-sight projection of
the orbital velocity of the star. We calculated the model flux during
transit using the linear limb-darkening law $B_{\lambda}(\mu) = 1 -
u_{\lambda}(1-\mu)$, where $B_{\lambda}$ is the normalized stellar
surface brightness profile and $\mu$ is the cosine of the angle
between the normal to the stellar surface and the line of sight.  We
employed the ``small planet'' approximation as described by Mandel \&
Agol (2002).  We held the limb darkening parameter $u_{\lambda}$ fixed
at a value appropriate for a star with the assumed properties, and for
the bandpass concerned, according to the models of Claret \&
Hauschildt (2003) and Claret (2004). These values were $u_g = 0.73$,
$u_r =0.61$, $u_V= 0.62$, and $u_{b+y} = 0.67$.

The goodness-of-fit parameter is
\begin{equation}
\chi^2 = \chi^2_v + \chi^2_f =
         \sum_{n=1}^{N_v} \left( \frac{v_O - v_C}{\sigma_v} \right)^2 +
         \sum_{n=1}^{N_f} \left( \frac{f_O - f_C}{\sigma_f} \right)^2,
\end{equation}
where $v_O$ and $v_C$ are the observed and calculated radial
velocities, of which there are $N_v=7$, and $f_O$ and $f_C$ are the
observed and calculated fluxes, of which there are $N_f=2310$.  Of the
flux measurements, 679 are our $g$-band measurements, 574 are our
$r$-band measurements, 237 are our $V$-band measurements, and 820 are
the $(b+y)$/2 measurements of Sato et al.\ (2005). We minimized
$\chi^2$ using an AMOEBA algorithm (Press et al.\ 1992).

The radial velocity uncertainties $\sigma_v$ were taken from Table~2
of Sato et al.\ (2005). To estimate the uncertainties in our
photometry, we performed the following procedure. We expect the two
dominant sources of uncertainty to be scintillation noise ($\sigma_S$)
and Poisson noise ($\sigma_P$). Young (1967) advocated an approximate
scaling law for the fractional error due to scintillation noise (see
also Dravins et al.\ 1998):
\begin{equation}
\sigma_S = 0.06\ X^{7/4}
\left( \frac{D}{{\rm 1~cm}} \right)^{-2/3}
\left( \frac{t_{\rm exp}}{{\rm 1~s}} \right)^{-1/2}
\exp \left( -\frac{h}{{\rm 8000~m}} \right),
\label{eq:scintillation}
\end{equation}
where $X$ is the airmass, $D$ is the diameter of the aperture, $t_{\rm
exp}$ is the exposure time, and $h$ is the altitude of the
observatory. For each of our 3 time series, we assumed that the noise
in each point obeyed
\begin{equation}
\sigma = \sqrt{\sigma_P^2 + (\beta \sigma_S)^2},
\end{equation}
where $\sigma_S$ was calculated with Eq.~\ref{eq:scintillation}.  We
determined the constant $\beta$ by requiring $\chi^2/N_{\rm DOF} = 1$
for that particular time series. Thus we did {\it not} attempt to use
the $\chi^2$ statistic to test the validity of the model; rather, we
assumed the model is correct, and sought the appropriate weight for
each data point. The results for $\beta$ ranged from 1.2 to 1.5. To
estimate the uncertainties in each of the three Sato et al.\ (2005)
time series (which were already corrected for airmass), we simply
assigned an airmass-independent error bar to all the points such that
$\chi^2/N_{\rm DOF} = 1$. The results agreed well with the RMS values
quoted in Table~5 of Sato et al.\ (2005).

After assigning the weight of each data point in this manner, we
analyzed all the data simultaneously and found the best-fitting
solution. This solution is overplotted on the data in Figure~1. Our
photometry, after correcting for the overall flux scale and for
airmass, is given in Table~1.  We estimated the uncertainties in the
model parameters using a Monte Carlo algorithm, in which the
optimization was performed on each of $7\times 10^4$ synthetic data
sets, and the distribution of best-fitting values was taken to be the
joint probability distribution of the parameters. Each synthetic data
set was created as follows:
\begin{enumerate}

\item We randomly drew $N_f=2310$ flux measurements $\{t_n, f_n\}$
from the real data set.  We drew these points with replacement,
i.e., we allowed for repetitions of the original data points.  This
procedure, recommended by Press et al.\ (1992), estimates the
probability distribution of the measurements using the measured data
values themselves, rather than the more traditional approach of
assuming a certain model for the underlying process.

\item This procedure was impractical for the radial velocity
measurements, because the number of data points is too small. Instead,
in each realization, we discarded a single radial velocity measurement
chosen at random. This was intended as a test of the robustness of the
results to single outliers. To each of the remaining $N_v=6$ radial
velocities, we added a random number drawn from a Gaussian
distribution, with zero mean and a standard deviation equal to the
quoted 1~$\sigma$ uncertainty.

\item To account for the uncertainty in the stellar properties, we
assigned a stellar mass by picking a random number from a Gaussian
distribution with mean $1.30~M_\odot$ and standard deviation
$0.10~M_\odot$. Likewise, for the stellar radius, we used a Gaussian
distribution with a mean of $1.45~R_\odot$ and a standard deviation of
$0.10~R_\odot$. We note that this procedure does not take into account
the intrinsic correlation between these two variables that is expected
from models of stellar structure and evolution. Assuming a stellar
mass-radius relation would provide an independent constraint that
would reduce the overall uncertainty on the planetary radius (Cody \&
Sasselov 2002). We elected not to make such an assumption because the
stellar age, and therefore its evolutionary state, are not known with
sufficient precision.

\end{enumerate}

\section{Discussion and Conclusions}

\begin{figure}[p]
\epsscale{1.0} \plotone{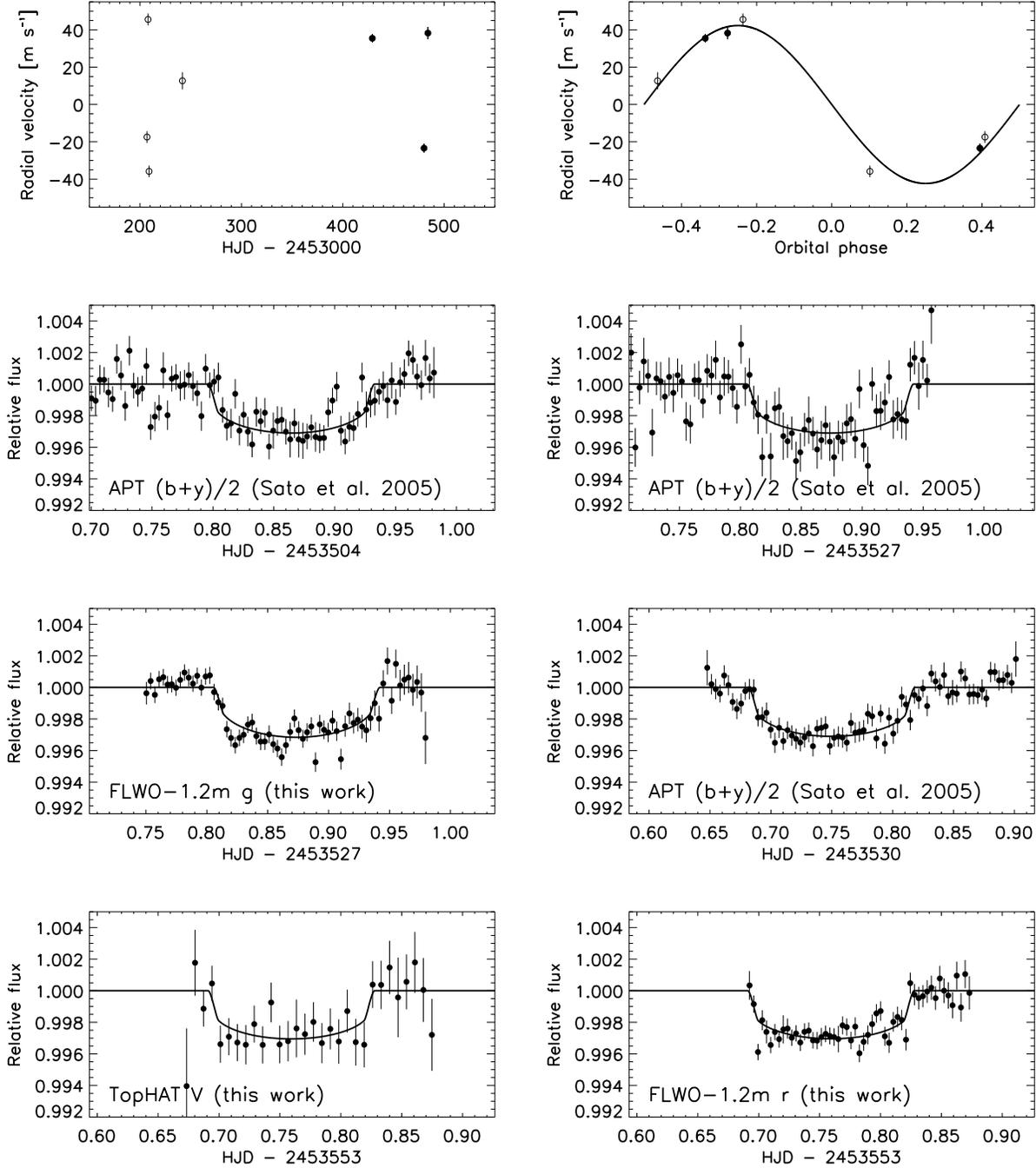}
\caption{Radial velocity variations and photometry of HD~149026.  The
upper two panels show the radial velocity measurements by Sato et al.\
(2005), as a function of time (left) and orbital phase (right).  Open
symbols are Subaru/HDS measurements, and filled circles are Keck/HIRES
measurements. The best-fitting value of $\gamma$ (for each
spectrograph) has been subtracted from the data. The best-fitting
model is overplotted. The remaining panels show the photometry from
this work and Sato et al.\ (2005). Although the points have been
averaged into 5 minute bins for presentation purposes (10 minute bins,
for TopHAT), the fitting procedure was performed on unbinned data.
\label{fig:phot}}
\end{figure}

The best-fitting model is illustrated in Fig.~1. The estimated
probability distribution for each model parameter is shown in Fig.~2.
Some of the parameters have correlated uncertainties, as shown in
Fig.~3. In addition to showing correlations among the model
parameters, Fig.~3 shows the distributions for three fundamental
properties of a single-transit light curve: the transit depth in the 
absence of stellar limb-darkening, defined
as $(R_{\rm P}/R_{\rm S})^2$; the transit duration, defined as the
time between first and fourth contact ($t_{\rm IV} - t_{\rm I}$); and
the ingress duration, defined as the time between first and second
contact ($t_{\rm II} - t_{\rm I}$). Since we have assumed a circular
Keplerian orbit, the ingress and egress durations are equal. The
contact times can be calculated from our model parameters via the
relations
\begin{eqnarray}
\sin i \hspace{0.02in} \cos \frac{\pi (t_{\rm IV} - t_{\rm I})}{P} & = & 
\sqrt{1 - \left(\frac{R_{\rm S} + R_{\rm P}}{a}\right)^2} \nonumber \\
\sin i \hspace{0.02in} \cos \frac{\pi (t_{\rm II} - t_{\rm I})}{P} & = & 
\sqrt{1 - \left(\frac{R_{\rm S} - R_{\rm P}}{a}\right)^2},
\end{eqnarray}
where $a$ is the semi-major axis, given by Kepler's law $G(M_{\rm S} +
M_{\rm P})/a^3 = (2\pi/P)^2$.

\begin{figure}[p]
\epsscale{1.0} \plotone{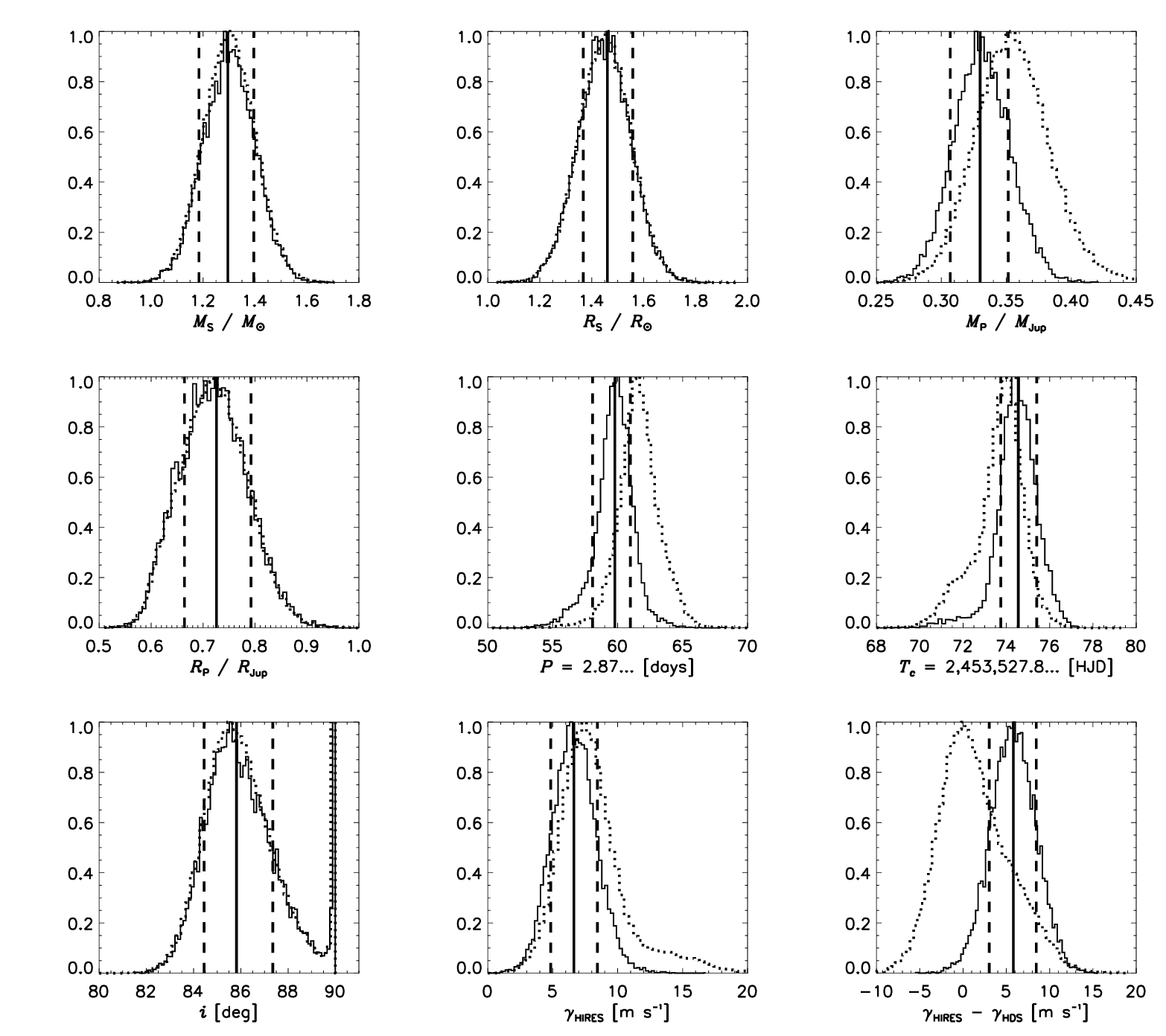}
\caption{ Estimated probability distributions of some planetary,
stellar, and orbital parameters. The dotted histograms show the
results of fits to all $7\times 10^4$ Monte Carlo realizations of the
data. The solid histograms show the results for the cases when the
biggest radial velocity outlier is dropped; we favor these results for
reasons described in the text. The one-dimensional
probability distribution for $i$ (lower left panel) shows a secondary peak 
near $i = 90\arcdeg$, which contains 9\% of the solutions.
We clip these solutions prior to determining the best-fit values and confidence limits
(see text).  Solid vertical lines show the median
value of each of the solid histograms. Dashed vertical lines show the
68.3\% confidence limits.
\label{fig:probdist}}
\end{figure}

\begin{figure}[p]
\epsscale{1.0} \plotone{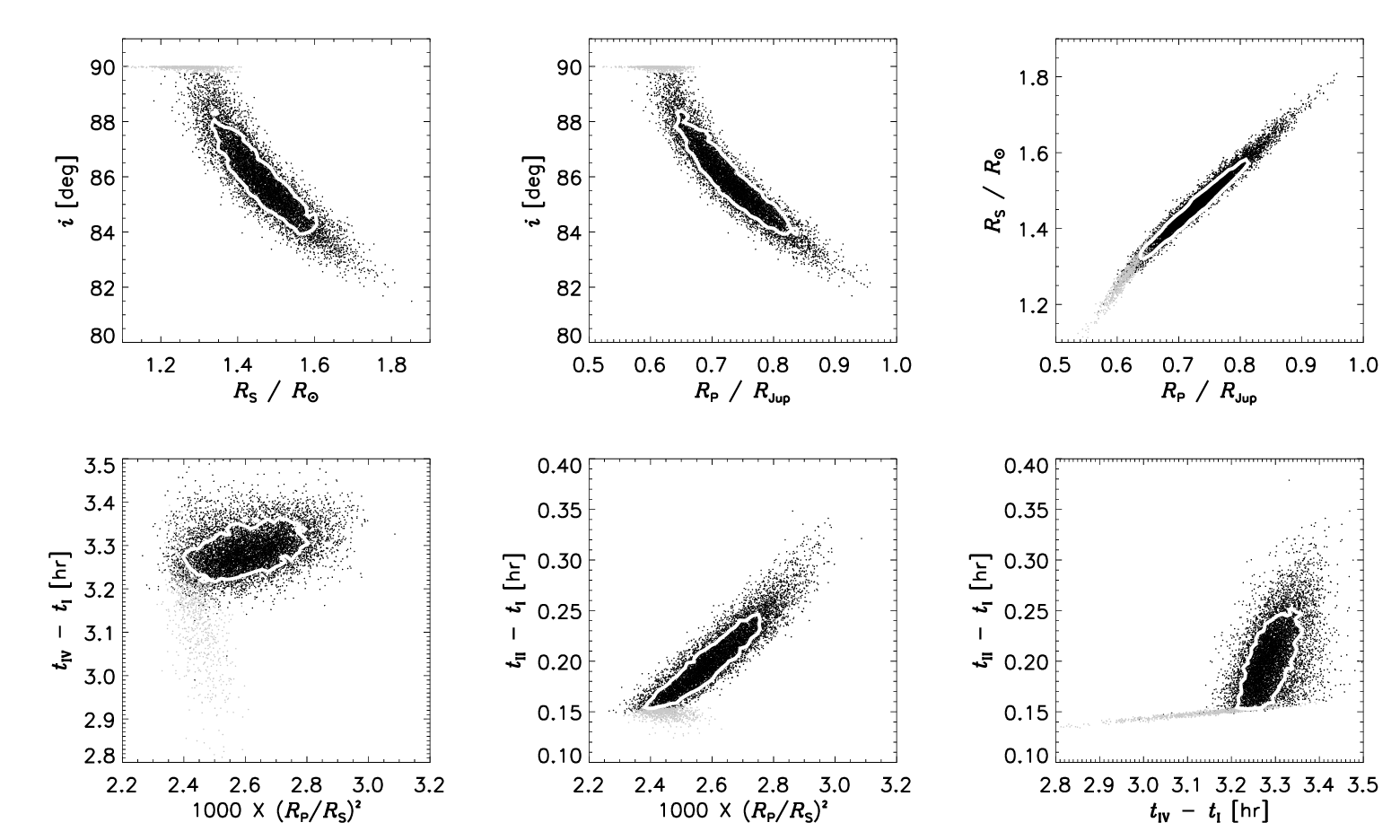}
\caption{ Joint probability distributions of some planetary, stellar,
and orbital parameters, based on the $10^4$ Monte Carlo realizations
in which $v_3$ was dropped. The density of points is proportional to
the probability density. The white contours are isoprobability
contours enclosing 68.3\% of the points. Solutions with $i>89\fdg75$
(an arbitrary threshold) are colored gray to distinguish between the
two solution branches described in the text.
\label{fig:correlations}}
\end{figure} 

We found that the probability distributions did not depend
significantly on which of the 7 radial velocity measurements were
discarded, with the important exception of the third Subaru/HDS
measurement ($v_3 = -42.48 \pm 3.10$~m~s$^{-1}$ at
HJD~2453208.909939). When this point was dropped, the results for the
orbital period, planetary mass, and velocity offsets changed by
1~$\sigma$. In Fig.~2, the solid histograms show the results of the
$10^4$ trials for which $v_3$ was discarded, and the dotted histograms
show the results of all $7\times 10^4$ trials. We believe that either
the model fails to accurately describe $v_3$ within its quoted
uncertainty (due to a missing ingredient such as a nonzero
eccentricity\footnote{We verified that a nonzero eccentricity is
sufficient to account for $v_3$ within the quoted uncertainty. In the
best-fitting model, $e=0.1$ and $\chi^2_v = 0.8$.} or an additional
planetary companion), or that the true uncertainty in this measurement
is larger than the estimate (due to stellar jitter or some other
source of systematic error). In support of this claim, we note that
this single point makes a disproportionate contribution to $\chi^2_v$.
When any single radial velocity measurement besides $v_3$ is dropped,
the minimum $\chi^2_v$ ranges from 13 to 19. Yet when $v_3$ is
dropped, $\chi^2_v = 0.5$. 
In Table~2, we present the results for the
cases in which $v_3$ is dropped. Our intention is to avoid biased
results from an oversimplified model or an underestimated uncertainty.

The one-dimensional probability distribution for $i$ shows that the
majority of solutions favor $i\approx 86\arcdeg$, but a peak is
evident at $i=90\arcdeg$ (see the lower left panel of Fig.~2). We
identify these maximum-$i$ solutions by applying an arbitrary cut-off
of $i > 89\fdg75$, and we display these solutions with a distinct
coloring in Fig.~3. These solutions, which account for 9\% of the total,
represent a pile-up at equatorial
configurations owing to the numerical constraint $i \le 90\arcdeg$.
We exclude this solution subset prior to determining the
best-fit values and confidence limits listed in Table~2.
We note that the duration of ingress and egress is significantly
shorter for HD~149026 than for either HD~209458 or TrES-1. The
available photometry samples the times of ingress and egress only
sparsely. We encourage high-cadence monitoring of these key portions
of the transit curve.

Based on this analysis, our best estimate of the transit ephemeris is
\begin{equation}
T_c~[{\rm HJD}] = (2453527.87455^{+0.00085}_{-0.00091}) + (2.87598^{+0.00012}_{-0.00017}) \ N.
\end{equation}

Tabulation of the observed time of each eclipse is of interest because
deviations from the predictions of a single-period ephemeris could
indicate the presence of planetary satellites or additional planetary
companions (Brown et al.\ 2001, Miralda-Escud\'{e} 2002). In
particular, terrestrial planets that induce a radial-velocity
perturbation below current detection limits can nonetheless be
detected through accurate eclipse timing (Holman \& Murray 2005, Agol
et al.\ 2005).  We searched for evidence of timing anomalies as
follows.  We constructed a model transit light curve for each of the 4
bandpasses [$g$, $r$, $V$, and $(b+y)/2$], based on the optimal
parameters appearing in Table~2 and the appropriate limb darkening
coefficient (\S~3). For each of the 6 extinction-corrected, normalized
light curves (identified by an index $j$), we then evaluated the
$\chi^2$ of the fit as a function of assumed time of center of
eclipse, $T_{c}^{j}$.  After we identified the optimal value for
$T_{c}^{j}$, we assigned uncertainties by identifying the timing
offsets at which the value of $\chi^2$ had increased by 1.  We list
these values in Table~3, and plot the ``observed minus calculated''
($O-C$) residuals in Figure~4.  The typical timing precision is
2~minutes, which is comparable to results from other ground-based
photometry (for a tabulation of $O-C$ for other transiting-planet
systems, see Charbonneau et al.\ 2005 for TrES-1, and Wittenmyer et
al.\ 2005 for HD~209458).  We find no evidence for deviations from the
predictions of a constant orbital period. We encourage future
monitoring of the times of eclipse of this system.  In particular,
space-based observations (Brown et al.\ 2001) should achieve a
substantial improvement in timing precision, and thus permit a more
sensitive search for perturbing planets.

\begin{figure}[p]
\epsscale{0.8} \plotone{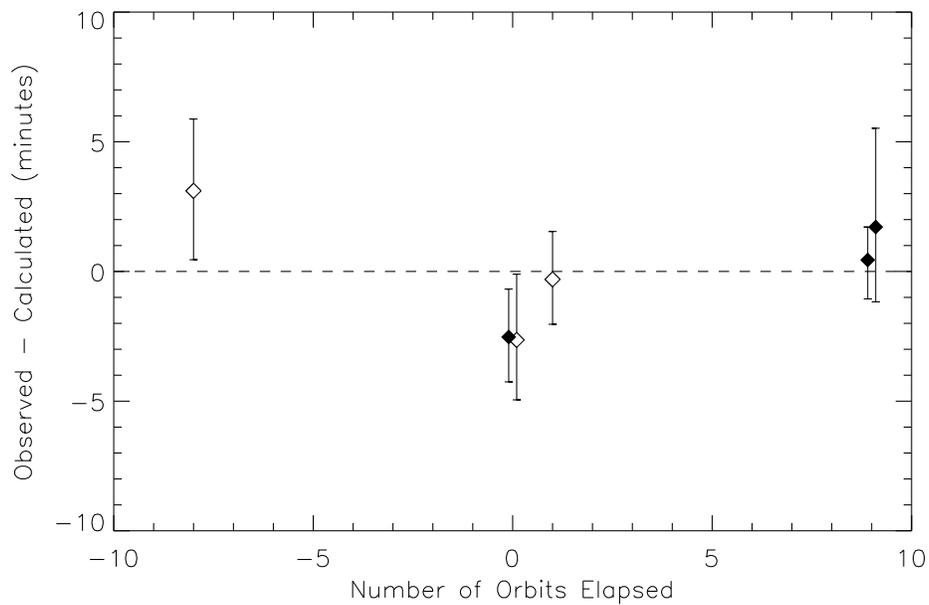}
\caption{The ``observed minus calculated'' ($O-C$) residuals for the 3 light curves
presented in this paper (black diamonds) and for the 3 light curves
in Sato et al.\ (2005, open diamonds).  In the two cases where two estimates are
available for the same transit, we have slightly offset the values along the horizontal axis for clarity.
The assumed ephemeris is given in Eq.~5.
The data plotted here are given in Table~3. 
\label{fig:ocresid}}
\end{figure}

Due to its favorable apparent brightness, the HD~149026 system will be
particularly amenable to a variety of follow-up studies.  Such
pursuits will be more observationally challenging than was the case
for HD~209458, owing to the smaller size of the planet relative to the
parent star.  Nonetheless, we are certain that such challenges will be
met and overcome.  Now that a handful of transiting planets of bright
stars have been identified, the long-sought-after goal of comparative
exoplanetology may be realized.

\acknowledgments We are very grateful to B.\ Sato, D.\ Fischer, G.\
Laughlin, and the other members of the N2K consortium, for sharing
their results in advance of publication. We thank T.\ Spahr, R.\ Kirshner, and 
M.\ Hicken for swapping telescope nights on short notice,
and H.\ Knutson for assistance with the remote observations.
Some of the Minicam observations on the FLWO 48-inch telescope
were obtained and reduced with support from the Kepler Mission
to SAO and PSI.  The TopHAT observations were supported
by NASA grant NNG04GN74G.
Work by J.N.W.\ was supported by NASA through grant HST-HF-01180.02-A,
and work by G.B.\ was supported by NASA through grant HST-HF-01170.01-A,
awarded by the Space Telescope Science Institute, which is operated by 
the Association of Universities for Research in Astronomy, Inc., for NASA, 
under contract NAS 5-26555.

\clearpage

\begin{deluxetable}{lcccc}
\tabletypesize{\normalsize}
\tablecaption{Photometry of HD~149026\label{tbl:photometry}}
\tablewidth{0pt}

\tablehead{
\colhead{Telescope} & \colhead{Filter} & \colhead{HJD} & \colhead{Relative flux} & \colhead{Uncertainty} \\
}

\startdata
    FLWO48 &   g & $      2453527.750311$ & $         0.9983$ & $         0.0018$ \\
    FLWO48 &   g & $      2453527.750624$ & $         0.9984$ & $         0.0018$ \\
    FLWO48 &   g & $      2453527.750936$ & $         0.9988$ & $         0.0018$ \\
    FLWO48 &   g & $      2453527.751260$ & $         0.9999$ & $         0.0018$ \\
    FLWO48 &   g & $      2453527.751585$ & $         1.0006$ & $         0.0018$
\enddata

\tablecomments{The quoted uncertainties are based on the procedure
described in \S~2, which assumes that our model is correct. We intend
for this Table to appear in entirety in the electronic version of the
Astronomical Journal. A portion is shown here to illustrate its
format. The data are also available in digital from the authors upon
request.}

\end{deluxetable}

\begin{deluxetable}{llll}

\tabletypesize{\normalsize}
\tablecaption{System Parameters of HD~149026\label{tbl:params}}
\tablewidth{0pt}

\tablehead{
\colhead{Parameter} & \colhead{Best fit} & \multicolumn{2}{c}{68\% conf.\ limits} \\
\colhead{ }         & \colhead{ }        & \colhead{lower} & \colhead{upper}
}

\startdata
                                     $M_{\rm S}$~[$M_\odot$]& $           1.30$ & $          -0.10$ & $ +           0.10$ \\
                                     $R_{\rm S}$~[$R_\odot$]& $           1.45$ & $          -0.10$ & $ +           0.10$ \\
                                 $M_{\rm P}$~[$M_{\rm Jup}$]& $          0.330$ & $         -0.023$ & $ +          0.022$ \\
                                 $R_{\rm P}$~[$R_{\rm Jup}$]& $          0.726$ & $         -0.062$ & $ +          0.066$ \\
                                                  $P$~[days]& $        2.87598$ & $       -0.00017$ & $ +        0.00012$ \\
                                       $T_c - 2453527$~[HJD]& $         0.87455$ & $       -0.00091$ & $ +        0.00085$ \\
                                                   $i$~[deg]& $           85.8$ & $           -1.3$ & $ +            1.6$ \\
                           $\gamma_{\rm HIRES}$~[m~s$^{-1}$]& $            6.6$ & $           -1.8$ & $ +            1.8$ \\
        $\gamma_{\rm HIRES} - \gamma_{\rm HDS}$~[m~s$^{-1}$]& $            5.8$ & $           -2.8$ & $ +            2.6$
\enddata

\tablecomments{Results are based on fits to the $10^4$ Monte Carlo
realizations of the data in which $v_3$ was dropped (see \S~3). Values
for $M_{\rm S}/M_\odot$ were drawn from a Gaussian random distribution
with mean 1.30 and standard deviation 0.1. Values for $R_{\rm
S}/R_\odot$ were drawn from a Gaussian random distribution with mean
1.45 and standard deviation 0.1. All other quantities listed in this
table were free parameters in the model.}

\end{deluxetable}

\begin{deluxetable}{llllll}
%\tabletypesize{\small}
%\rotate
\tablecaption{Observed Times of Transit}
\tablewidth{0pt}
\tablehead{
\colhead{Event} & \colhead{$T_{c} \ [{\rm HJD}]$} & \colhead{$N_{\rm elapsed}$} & \colhead{$\sigma_{\rm HJD}$} & \colhead{$(O-C)$} & \colhead{$\frac{(O-C)}{\sigma_{\rm HJD}}$}}
\startdata
    Sato et al.\ (2005)    &  2453504.8689 & $-$8.0           &  0.0019 &   $+$0.0022  &  $+$1.15\\
    Sato et al.\ (2005)    &  2453527.8727 & $\phantom{+}$0.0 &  0.0017 &   $-$0.0018  &  $-$1.09\\
    This work [FLWO48 $g$] &  2453527.8728 & $\phantom{+}$0.0 &  0.0012 &   $-$0.0018  &  $-$1.41\\
    Sato et al.\ (2005)    &  2453530.7503 & $+$1.0           &  0.0012 &   $-$0.0002  &  $-$0.17\\
    This work [FLWO48 $r$] &  2453553.7587 & $+$9.0           &  0.0010 &   $+$0.0003  &  $+$0.32\\
    This work [TopHAT $V$] &  2453553.7597 & $+$9.0           &  0.0023 &   $+$0.0012  &  $+$0.51
\enddata

\end{deluxetable}

\end{document}